\documentclass[runningheads,envcountsect,envcountsame]{llncs}

\usepackage[T1]{fontenc}
\usepackage{lmodern,amsmath,amssymb,mathtools,microtype}
\usepackage{array,tabularx,float,booktabs}
\usepackage[hidelinks]{hyperref}

\floatstyle{ruled}
\newfloat{algorithm}{htbp}{loa}
\floatname{algorithm}{Algorithm}

\newcommand{\E}{\mathbb E}
\newcommand{\OPT}{\operatorname{OPT}}
\newcommand{\rank}{\operatorname{rk}}
\newcommand{\WF}{\textnormal{\textsc{Waterfill}}}
\newcommand{\LA}{\textnormal{\textsc{Lookahead}}}
\newcommand{\waste}{\Phi}
\newcolumntype{Y}{>{\raggedright\arraybackslash}X}

\title{Randomized MST Certification beyond Cacti}
\titlerunning{Randomized MST Certification beyond Cacti}

\author{Sina Ghasemi Nezhad\inst{1,2}}
\authorrunning{S. Ghasemi Nezhad}

\institute{Department of Computer Science, Vrije Universiteit Amsterdam, The Netherlands\and
Informatics Institute, Faculty of Science, University of Amsterdam, The Netherlands\\
\email{sina.ghasemi.nezhad@student.uva.nl}}

\begin{document}

\maketitle

\begin{abstract}
We study minimum spanning tree certification under explorable edge uncertainty. Edge weights lie in open intervals or singleton sets, and querying an edge reveals its weight at a known positive cost. On general graphs the best known randomized competitive ratio is $1+1/\sqrt2$, while the optimal ratio $3/2$ is known only for cacti. We go beyond cacti in two ways, with conditions that refer to the uncertain edges outside the upper-limit tree rather than to the graph alone. For unit query costs we give a $3/2$-competitive algorithm, which is best possible, whenever every block has at most two such edges; this includes all graphs whose blocks are cycles or theta graphs. The algorithm replaces the constant budget of the water-filling method by budgets chosen with lookahead to the next fundamental cycle. For arbitrary positive query costs and every fixed $k\ge1$, we obtain ratio $1+\sqrt{k/(2(k+1))}$ whenever every block has at most $k$ such edges, in particular when every block has cycle rank at most $k$. This is below $1+1/\sqrt2$ for every $k$, and we show that it is the best ratio attainable with a constant water-filling budget.
\keywords{Explorable uncertainty \and Minimum spanning tree \and Randomized algorithm \and Competitive analysis \and Matroid.}
\end{abstract}

\section{Introduction}
In the minimum spanning tree problem under explorable uncertainty, each edge weight is initially specified by an interval. An algorithm adaptively queries edges until it can certify a minimum spanning tree (MST). The objective is to minimize query cost relative to the cheapest certificate for the same realized weights.

Erlebach et al.~\cite{erlebach2008computing} established the optimal deterministic competitive ratio of two. Megow, Mei\ss ner, and Skutella~\cite{megow2017randomization} showed that no randomized algorithm is better than $3/2$-competitive and obtained a randomized ratio of $1+1/\sqrt2\approx1.707$, including arbitrary positive query costs and minimum-weight matroid bases. Their algorithm processes the non-tree edges of a tentative tree one by one and uses a water-filling rule with a constant budget to decide, for each of them, which side of its fundamental cycle to query. Focke, Megow, and Mei\ss ner~\cite{focke2020minimum} studied preprocessing and implementations, and Erlebach et al.~\cite{erlebach2022learning} studied queries guided by predictions. For cactus graphs, Mathwieser and \c{C}ela~\cite{mathwieser2024special} proved the optimal randomized ratio of $3/2$. Closing the gap between $3/2$ and $1+1/\sqrt2$ on general graphs is open.

The cactus result exploits the separation of cycles: their edge sets are disjoint, and their certificates combine independently. Overlapping cycles interact because one query may contribute to several cycle decisions, and because resolving one cycle may change the tree and thereby the next cycle. We study the simplest classes with such interactions.

A \emph{block} of a connected graph is a maximal connected subgraph without a cut vertex; it is either a bridge or a maximal $2$-connected subgraph, and the blocks partition the edge set. The \emph{cycle rank} of a connected graph $H$ is $\mu(H)=|E(H)|-|V(H)|+1$, the number of edges outside any spanning tree of $H$. Let $\mathcal C_k$ be the class of connected simple graphs whose blocks have cycle rank at most $k$. Thus $\mathcal C_1$ is the class of cacti, and $\mathcal C_2$ consists of the graphs whose blocks are bridges, cycles, or theta graphs, where a \emph{theta graph} is the union of three internally disjoint paths with common distinct endpoints. The number of blocks, the path lengths, and the total cycle rank are unrestricted.

Our conditions are more general and depend on the uncertainty pattern. For a block $H$, let $u(H)$ be the number of uncertain edges of $H$ outside its upper-limit tree, the minimum spanning tree for weights just below the upper interval endpoints, and let $u^*(H)$ be the number of uncertain edges inside its lower-limit tree; the precise definitions are given in Section~\ref{sec:algorithm}. Always $u(H)\le\mu(H)$ and $u^*(H)\le|V(H)|-1$, but both numbers can be small in arbitrarily dense blocks.

\begin{theorem}\label{thm:opt}
For unit query costs, MST certification under explorable edge uncertainty admits a polynomial-time randomized $3/2$-competitive algorithm on all instances in which every block $H$ satisfies $\min\{u(H),u^*(H)\}\le2$. In particular, this holds on $\mathcal C_2$. The ratio is best possible.
\end{theorem}

\begin{theorem}\label{thm:main}
For arbitrary positive query costs and every integer $k\ge1$, MST certification under explorable edge uncertainty admits a polynomial-time randomized algorithm with competitive ratio
\begin{equation}\label{eq:rho}
 \rho_k=1+\sqrt{\frac{k}{2(k+1)}}
\end{equation}
on all instances in which every block $H$ satisfies $\min\{u(H),u^*(H)\}\le k$. In particular, this holds on $\mathcal C_k$ and whenever
$\min\{|V(H)|-1,\mu(H)\}\le k$ for every block.
\end{theorem}

Theorem~\ref{thm:opt} extends the optimal ratio from cacti to instances with overlapping cycles. In Theorem~\ref{thm:main}, $\rho_1=3/2$, $\rho_2=1+1/\sqrt3\approx1.577$, and $\rho_k<1+1/\sqrt2$ for every $k$; the limit for $k\to\infty$ is $1+1/\sqrt2$, so the theorem gives no uniform improvement on unrestricted instances. The parameter can be refined further to an \emph{overlap number} (Theorem~\ref{thm:corank}).

\paragraph{Techniques.}
Both results use the ordered framework of~\cite{megow2017randomization}, which we describe in a self-contained way in Section~\ref{sec:algorithm}. Its analysis charges query probabilities to potentials on the edges of an initial tree. For Theorem~\ref{thm:main} we keep the number $h$ of non-tree edges that can charge one tree edge. This replaces the integral bound $1/(2\alpha)$ of~\cite{megow2017randomization} by $h/(2\alpha(h+1))$, and balancing with the budget $\alpha$ yields~\eqref{eq:rho}. Duality permits using the rank instead of the cycle rank. We also show that this is exactly the worst-case ratio of water-filling with a constant budget, and that with unit costs no constant budget achieves less than $11/7$ on theta graphs (Section~\ref{app:tight}). Theorem~\ref{thm:opt} therefore needs a new ingredient. With two uncertain non-tree edges, the second fundamental cycle is known in advance unless the first step changes the tree. We prove that such a change forces two specific queries into every certificate, and we choose the first budget with lookahead to the unchanged second cycle. The first budget grows with the number of elements shared by the two cycles, and the second budget shrinks accordingly.

\section{Model}\label{sec:model}
Let $G=(V,E)$ be a connected simple graph. Each edge has a known query cost $c_e>0$ and an uncertainty set $I_e$, either an open bounded interval $(L_e,U_e)$ or a singleton $\{w_e\}$, in which case $L_e=U_e=w_e$. An edge is \emph{uncertain} in the first case and \emph{trivial} in the second. A realization $w\in\prod_e I_e$ is fixed before the algorithm's random choices. Querying an uncertain edge reveals $w_e$ exactly; trivial edges require no query. Write $c(Q)=\sum_{e\in Q}c_e$.

A set $Q$ of uncertain edges is a \emph{certificate} for $w$ if some spanning tree is minimum for every completion consistent with the revealed weights on $Q$. Let $\OPT(G,w)$ be the minimum certificate cost. An algorithm is $\rho$-competitive if
$\E[c(Q_{\rm alg})]\le\rho\,\OPT(G,w)$
for every instance and realization. The expectation concerns only the algorithm's random choices; weights have no assumed probability distribution.

\paragraph{Matroids.}
We also consider minimum-weight bases of a matroid $M=(E,\mathcal I)$ given by an independence oracle; certificates are defined by replacing trees with bases. Its rank is $\rank(M)$ and its corank is $|E|-\rank(M)$. For a basis $B$ and $f\notin B$, the \emph{fundamental circuit} $C(B,f)$ is the unique circuit in $B\cup\{f\}$; the set $(B\cup\{f\})\setminus\{x\}$ is a basis exactly when $x\in C(B,f)$. A basis $B$ has minimum weight if and only if $w_x\le w_f$ for all $f\notin B$ and $x\in C(B,f)$. If an element is the unique heaviest element of some circuit, it lies in no minimum-weight basis. For the graphic matroid $M(G)$, bases are spanning trees, circuits are cycles, and rank and corank are $|V|-1$ and $\mu(G)$. The dual matroid $M^*$ has the complements of bases as bases.

\begin{lemma}\label{lem:decomp}
If $\mathcal H$ is the set of blocks of $G$, then
\[
 \OPT(G,w)=\sum_{H\in\mathcal H}\OPT(H,w|_{E(H)}),
\]
and a set is a certificate for $G$ if and only if its restriction to every block is a certificate for that block.
For a matroid $M$, replacing it by $M^*$ and replacing every weight and uncertainty set by its negative preserves the family of certificates and their costs.
\end{lemma}
\begin{proof}
A spanning tree is minimum in $G$ exactly when its restriction to every block is minimum. A global certificate therefore restricts to a certificate in each block: otherwise a completion changing only that block would invalidate it. Conversely, the union of block certificates is a global certificate. Block edge sets are disjoint, giving the equality.

For every basis $B$ of $M$, its complement is a basis of $M^*$ and
$\sum_{e\in E\setminus B}(-w_e)=\sum_{e\in B}w_e-\sum_{e\in E}w_e$.
The final term is independent of $B$. Optimality of $B$ is thus equivalent to optimality of its complement under negated weights, simultaneously for every consistent completion. Query costs are unchanged.
\qed
\end{proof}

The description of $\mathcal C_2$ follows by suppressing the vertices of degree two in a $2$-connected block of cycle rank two: since $\sum_v(\deg(v)-2)=2$, the result has either one vertex of degree four, which would be a cut vertex of the block, or two vertices of degree three and no loop, hence three parallel edges.

\section{Framework}\label{sec:algorithm}
This section describes the framework of~\cite{megow2017randomization} for a matroid, so that a graph and its dual use the same procedure. All algorithms in this paper are instances of Algorithm~\ref{alg:wf}; they differ only in the \emph{level rule} used in its randomized step.

\paragraph{Preprocessing.}
A \emph{lower-limit basis} $B_L$ is a minimum basis when every uncertain weight is infinitesimally above its lower endpoint; an \emph{upper-limit basis} $B_U$ uses weights infinitesimally below the upper endpoint. Greedy computation orders elements by $(L_e,\mathbf1_{L_e<U_e},\operatorname{id}(e))$ and $(U_e,-\mathbf1_{L_e<U_e},\operatorname{id}(e))$, respectively, for fixed labels $\operatorname{id}$. Every uncertain element of $B_L\setminus B_U$ belongs to every certificate~\cite[Sections~3 and~8]{megow2017randomization}. We query these elements, treat queried elements as trivial, and recompute until the limit bases coincide. Let $P$ be the set of elements queried in this way and $A$ the common basis. The instance at this point is the \emph{residual instance}; its endpoints $L_e,U_e$ are \emph{frozen} and used in all definitions below. Since $P$ is part of every certificate, the certificates of the original instance are the sets $P\cup Q$ with $Q$ a certificate of the residual instance.

By the greedy rule, for every $f\notin A$ and every $x\in C(A,f)\setminus\{f\}$,
\begin{equation}\label{eq:limits}
 L_x\le L_f\quad\text{and}\quad U_x\le U_f .
\end{equation}

For an instance $(M,I)$ let $u(M,I)$ be the number of uncertain elements outside its upper-limit basis. For a block $H$ of a graph we put $u(H)=u(M(H),I)$ and $u^*(H)=u(M(H)^*,-I)$. The upper-limit basis of the dual instance is the complement of a lower-limit tree of $H$, so $u^*(H)$ counts the uncertain edges inside that tree.

\begin{lemma}\label{lem:uncertain}
Querying an element does not increase $u(M,I)$. Consequently, the residual instance has at most $u(M,I)$ uncertain elements outside $A$.
\end{lemma}
\begin{proof}
Let $B$ be the upper-limit basis and let $e$ be an uncertain element. Querying $e$ moves it to an earlier position of the greedy order and leaves the relative order of all other elements unchanged. Let $B'$ be the new greedy basis. An element $z\ne e$ outside $B$ is spanned by its predecessors in the old order; its set of predecessors can only gain $e$, so $z\notin B'$. Hence $B'\subseteq B\cup\{e\}$, and since $|B'|=|B|$, either $B'=B$ or $B'=(B\cup\{e\})\setminus\{x\}$ for one element $x$. If $e\in B$, then $e$ stays in the basis and $B'=B$. If $e\notin B$, then $e$ is no longer counted, and at most the element $x$ is newly counted. In the residual instance, $A$ is the upper-limit basis.
\qed
\end{proof}

\paragraph{Arrivals.}
The elements of $F=E\setminus A$ are called \emph{arrivals} and are processed in increasing order of $(L_f,U_f,\operatorname{id}(f))$. By Lemma~\ref{lem:uncertain}, at most $u(M,I)$ of them are uncertain. The algorithm maintains a working basis $D$, initially $A$, that is a certified minimum basis of the elements seen so far. When $f$ arrives, let $C_f=C(D,f)$. The algorithm queries until the current information certifies a maximum of $C_f$, that is, an element $g\in C_f$ such that every other element of $C_f$ has supremum at most the infimum of $g$. A tie for the maximum then involves only elements with known weight. The \emph{canonical maximum} is $f$ if $f$ attains the maximum, and otherwise the maximum-weight element with the largest label. It is deleted from $D\cup\{f\}$.

Which elements must be queried at $f$? The \emph{witness set} of $f$ is
\begin{equation}\label{eq:witness}
 X_f=\{a\in A\cap C_f: a\text{ is uncertain and }U_a>L_f\},
\end{equation}
the initial basis elements on the circuit whose intervals reach above the lower limit of $f$. As long as neither $f$ nor some $a\in X_f$ is known, either of them may be the heavier one, and no certificate is possible. The algorithm therefore starts with a \emph{core step}: it queries either $f$ or all of $X_f$. Afterwards, in the \emph{completion step}, it repeatedly queries an unknown element of $C_f$ with the largest current upper endpoint until a maximum is certified. Only the core step involves a choice, and only this choice can be wasteful.

\begin{example}\label{ex:triangle}
Consider a triangle with $I_b=\{1\}$, $I_a=(0,3)$, and $I_f=(1,4)$. Both limit trees equal $A=\{a,b\}$, the only arrival is $f$, and $X_f=\{a\}$. If $w_a=2$ and $w_f=7/2$, then $\{f\}$ is a certificate, whereas querying $a$ first reveals a value inside $I_f$ and the completion step must still query $f$. If $w_a=1/2$ and $w_f=2$, then $\{a\}$ is a certificate, whereas querying $f$ first forces a query of $a$. A core step that queries $f$ or $a$ with probability $1/2$ each has expected cost $3/2$ in both cases, which matches the lower bound of~\cite{megow2017randomization}.
\end{example}

\begin{lemma}\label{lem:framework}
Consider the residual instance and an arrival $f$.
\begin{enumerate}
\item[\textnormal{(i)}] \emph{Witness property.} Every certificate contains $f$ or all of $X_f$.
\item[\textnormal{(ii)}] \emph{Completion.} Every element queried in the completion step for $f$ belongs to every certificate.
\item[\textnormal{(iii)}] \emph{Determinism.} For a fixed realization, the working bases, the circuits $C_f$, the canonical maxima, and the sets $X_f$ do not depend on the queries chosen in core steps.
\end{enumerate}
\end{lemma}
\begin{proof}
Properties (i) and (ii) are the witness lemma of~\cite[Section~4]{megow2017randomization} and its two consequences for the cycle closed by an arriving edge, with the matroid extension in Section~8 of that work. For (iii), induct on the arrivals. The initial basis and the arrival order are fixed. The realization determines the canonical maximum of each circuit, and deleting it determines the next working basis. The frozen endpoints then determine $X_f$.
\qed
\end{proof}

\paragraph{Level rules.}
The algorithm draws a single threshold $\beta$ uniformly from $[0,1]$. A \emph{level rule} assigns to every arrival $f$ with $X_f\ne\varnothing$ a level $t_f\in[0,1]$ that may depend on the residual instance and on the canonical maxima of earlier arrivals, but not on $\beta$. The core step queries $f$ if $\beta>t_f$ and all of $X_f$ otherwise. A trivial arrival always receives level one: certificates consist of uncertain elements, so by the witness property its witness set is contained in every certificate, and querying it is never wasteful. By Lemma~\ref{lem:framework}(iii), the levels are fixed numbers for a fixed realization. The \emph{potential} of $a\in A$ is $y_a=\max\{t_f:a\in X_f\}$, or zero if $a$ lies in no witness set.

\begin{algorithm}[t]
\caption{Ordered framework with a level rule}\label{alg:wf}
\small
\begin{tabularx}{\linewidth}{@{}rY@{}}
\multicolumn{2}{@{}l}{\textbf{Input:} Matroid $M$, uncertainty sets $I$, costs $c$, and a level rule.}\\
\multicolumn{2}{@{}l}{\textbf{Output:} A certified minimum basis $D$ and the set $Q$ of queried elements.}\\[3pt]
1 & Preprocess to a common limit basis $A$ and query set $P$; freeze the endpoints.\\
2 & $D\gets A$; $Q\gets P$; draw $\beta$ uniformly from $[0,1]$.\\
3 & \textbf{for} $f\in E\setminus A$ in increasing order of $(L_f,U_f,\operatorname{id}(f))$ \textbf{do}\\
4 & \quad $C\gets C(D,f)$; \ $X\gets\{a\in A\cap C: a\text{ uncertain},\ U_a>L_f\}$.\\
5 & \quad \textbf{if} $X\ne\varnothing$ \textbf{then} \hfill (core step)\\
6 & \qquad obtain $t_f$ from the level rule; \textbf{if} $\beta>t_f$ query $f$ \textbf{else} query all of $X$.\\
7 & \quad \textbf{while} no maximum of $C$ is certified \textbf{do} \hfill (completion step)\\
8 & \qquad query an unknown element of $C$ with the largest current upper endpoint.\\
9 & \quad Let $m$ be the canonical maximum of $C$; \ $D\gets(D\cup\{f\})\setminus\{m\}$.\\
10 & \textbf{return} $(D,Q)$.
\end{tabularx}
\par\smallskip
A query of a known element has no effect. Every new query reveals one weight and adds its element to $Q$.
\end{algorithm}

We record one more consequence of~\eqref{eq:limits}. If a trivial element $g$ arrives while the working basis is still $A$, then every $x\in C(A,g)\setminus\{g\}$ satisfies $w_x\le U_x\le w_g$. Hence $g$ is the canonical maximum, this is certified without any query, $X_g=\varnothing$, and the working basis remains $A$. In particular, the working basis equals $A$ when the first uncertain arrival is processed.

\begin{lemma}\label{lem:account}
Let $Q^*$ be a certificate of the residual instance, and let $Q$ be the query set of Algorithm~\ref{alg:wf} with any level rule. Then
\begin{equation}\label{eq:account}
 \E[c(Q\setminus P)]\le c(Q^*)+\waste(Q^*),\qquad
 \waste(Q^*)=\sum_{a\in A\setminus Q^*}c_a y_a+
 \sum_{\substack{f\in F\setminus Q^*\\X_f\ne\varnothing}}c_f(1-t_f).
\end{equation}
Consequently, if $\waste(Q^*)\le\gamma\,c(Q^*)$ for every realization and every certificate $Q^*$ of the residual instance, then the algorithm is $(1+\gamma)$-competitive.
\end{lemma}
\begin{proof}
By Lemma~\ref{lem:framework}(ii), every query outside $Q^*$ is made in a core step. An element $a\in A\setminus Q^*$ is queried in a core step only if $\beta\le t_f$ for some arrival $f$ with $a\in X_f$, hence with probability at most $y_a$. An arrival $f\notin Q^*$ is queried in a core step only if $X_f\ne\varnothing$ and $\beta>t_f$, hence with probability at most $1-t_f$. Elements of $Q^*$ contribute at most $c(Q^*)$. For the last claim, apply~\eqref{eq:account} to a minimum certificate $Q^*$ of the residual instance; the original optimum is $c(P)+c(Q^*)$, and the algorithm pays $c(P)$ for preprocessing.
\qed
\end{proof}

\begin{proposition}\label{prop:correct}
Algorithm~\ref{alg:wf} returns a certified minimum basis, queries no element twice, and uses polynomially many independence-oracle calls and arithmetic operations, provided the level rule can be evaluated in polynomial time.
\end{proposition}
\begin{proof}
Inserting an arrival into a certified minimum basis and deleting a certified maximum of its circuit preserves a certified minimum basis of the processed elements. The while loop terminates because a maximum is certified once all elements of $C$ are known. Preprocessing and the fundamental circuits need $O(|E|^2)$ oracle calls.
\qed
\end{proof}

\paragraph{Application to graphs.}
By Lemma~\ref{lem:decomp}, blocks are processed independently; a bridge needs no query. For a block $H$ we run Algorithm~\ref{alg:wf} on the instance $(M(H),I)$, or on the dual instance $(M(H)^*,-I)$ and complement the returned basis, whichever has the smaller value of $u$. There are $\mu(H)$ arrivals in the first case and $|V(H)|-1$ in the second, so $u(H)\le\mu(H)$ and $u^*(H)\le|V(H)|-1$.

\section{Constant Budgets}\label{sec:finite}
The \emph{water-filling rule} of~\cite{megow2017randomization} with budget $\alpha>0$ keeps a potential $y_a$ for every $a\in A$, initially zero. For an uncertain arrival $f$ with $X_f\ne\varnothing$ it sets
\begin{equation}\label{eq:fill}
 t_f=\max\Bigl\{t\in[0,1]:
       \sum_{a\in X_f}c_a(t-y_a)_+\le\alpha c_f\Bigr\}
\end{equation}
and raises the potentials of $X_f$ to $\max\{y_a,t_f\}$; a trivial arrival has level one. This is a level rule, and the final potentials agree with the definition in Section~\ref{sec:algorithm}. We write \WF$(\alpha)$ for Algorithm~\ref{alg:wf} with this rule. The original analysis yields ratio $1+\max\{\alpha,1/(2\alpha)\}$. We refine it using the following parameter.

In the residual instance, an uncertain $a\in A$ and an uncertain arrival $f$ \emph{overlap} if $L_f<U_a$. The \emph{overlap number} is the largest number of arrivals overlapping a single element of $A$. It is at most the number of uncertain arrivals, hence at most $u(M,I)$ by Lemma~\ref{lem:uncertain}, and by~\eqref{eq:witness} every element of $X_f$ overlaps $f$ if $f$ is uncertain.

Fix a realization and a certificate $Q^*$ of the residual instance, and put $A_*=A\cap Q^*$ and $F_*=F\cap Q^*$.

\begin{lemma}\label{lem:first}
For every $\alpha>0$,
$\sum_{a\in A\setminus Q^*}c_a y_a\le\alpha\, c(F_*)$.
\end{lemma}
\begin{proof}
If an arrival $f$ raises the potential of some $a\notin Q^*$, then $X_f\not\subseteq Q^*$, so $f\in Q^*$ by the witness property, Lemma~\ref{lem:framework}(i). In particular $f$ is uncertain, and by~\eqref{eq:fill} the total cost-weighted increase caused by $f$ is at most $\alpha c_f$. Sum over these arrivals.
\qed
\end{proof}

\begin{lemma}\label{lem:finite}
If the overlap number is at most $h\ge1$, then
\begin{equation}\label{eq:second}
 \sum_{f\in F\setminus Q^*,\,X_f\ne\varnothing}c_f(1-t_f)
 \le\frac{h}{2\alpha(h+1)}\,c(A_*).
\end{equation}
\end{lemma}
\begin{proof}
Let $\Delta_{f,a}\ge0$ be the increase of $y_a$ at arrival $f$. When $t_f<1$, the budget in~\eqref{eq:fill} is exhausted:
$\sum_{a\in X_f}c_a\Delta_{f,a}=\alpha c_f$.
If also $f\notin Q^*$, then $X_f\subseteq A_*$ by the witness property. Terms with $t_f=1$ vanish; these include all trivial arrivals. Hence
\begin{equation}\label{eq:redistribute}
 \sum_{f\in F\setminus Q^*,\,X_f\ne\varnothing}c_f(1-t_f)
 =\sum_{a\in A_*}\frac{c_a}{\alpha}
    \sum_{\substack{f\in F\setminus Q^*\\t_f<1}}
        (1-t_f)\Delta_{f,a}.
\end{equation}
Fix $a\in A_*$ and list the arrivals of the inner sum with $\Delta_{f,a}>0$ in their order of arrival; let their increments be $d_1,\ldots,d_q$ and their levels $t_1,\ldots,t_q$. Each of them is uncertain, has $a$ in its witness set, and thus overlaps $a$, so $q\le h$. If $q=0$, the inner sum is zero. Otherwise put $s_i=d_1+\cdots+d_i$ and $D=s_q$. The key observation is that $t_i\ge s_i$: a potential is only ever raised to the current water level, so immediately after the $i$th listed arrival $y_a$ equals $t_i$; on the other hand $y_a$ starts at zero, never decreases, and has by then received at least the increments $d_1,\ldots,d_i$. Therefore
\begin{align}
 \sum_{i=1}^q(1-t_i)d_i
 &\le\sum_{i=1}^q(1-s_i)d_i
 =D-\frac{D^2}{2}-\frac12\sum_{i=1}^q d_i^2 \nonumber\\
 &\le D-\frac{q+1}{2q}D^2
 \le\frac{q}{2(q+1)}
 \le\frac{h}{2(h+1)}. \label{eq:quadratic}
\end{align}
The equality uses $\sum_i s_id_i=\frac12\bigl(D^2+\sum_i d_i^2\bigr)$. The second line uses $\sum_i d_i^2\ge D^2/q$ by the Cauchy--Schwarz inequality and maximizes the quadratic at $D=q/(q+1)$. Substitution in~\eqref{eq:redistribute} proves the claim.
\qed
\end{proof}

\begin{theorem}\label{thm:corank}
Consider an uncertainty matroid whose residual instance has overlap number at most $h\ge1$; in particular, $h$ may be $u(M,I)$ or the corank. Then \WF$(\alpha)$ has competitive ratio at most
\begin{equation}\label{eq:budgetbound}
 1+\max\left\{\alpha,\frac{h}{2\alpha(h+1)}\right\},
\end{equation}
which equals $\rho_h$ for $\alpha=\sqrt{h/(2(h+1))}$. With overlap number zero, every query lies in every certificate and the algorithm is optimal.
\end{theorem}
\begin{proof}
Lemmas~\ref{lem:first} and~\ref{lem:finite} give
$\waste(Q^*)\le\alpha\, c(F_*)+\frac{h}{2\alpha(h+1)}\,c(A_*)$.
Since $A_*$ and $F_*$ partition $Q^*$, Lemma~\ref{lem:account} applies with $\gamma$ the larger coefficient. The stated budget equalizes the coefficients. If no pair overlaps, the witness sets of all uncertain arrivals are empty, and all queries are in every certificate by Lemma~\ref{lem:framework}(i) and~(ii).
\qed
\end{proof}

The budget may be chosen after preprocessing, as a function of the overlap number of the residual instance. Thus \WF\ is $\rho_h$-competitive on arbitrary graphs whenever, after preprocessing, the interval of every tree edge reaches above the lower limits of at most $h$ uncertain non-tree edges of its block.

\begin{proof}[of Theorem~\ref{thm:main}]
For a block $H$, choose among $(M(H),I)$ and $(M(H)^*,-I)$ the instance with the smaller value $h\le k$ of $u$. By Lemma~\ref{lem:uncertain} its residual instance has overlap number at most $h$; run \WF\ with the budget of Theorem~\ref{thm:corank} for this overlap number. Since $\rho_h$ increases with $h$, every block is handled with ratio at most $\rho_k$, and Lemma~\ref{lem:decomp} combines the blocks and transfers the dual guarantee. The last statement follows from $u(H)\le\mu(H)$ and $u^*(H)\le|V(H)|-1$.
\qed
\end{proof}

The bound~\eqref{eq:budgetbound} is the exact worst-case ratio of \WF$(\alpha)$ for every $h$ and every $\alpha\le1$, so $\rho_h$ is the best guarantee a constant budget can give (Proposition~\ref{prop:tightcost}). With unit costs and two arrivals, no constant budget achieves a ratio below $11/7$ (Proposition~\ref{prop:tight}). The next section overcomes this barrier.

\section{Lookahead}\label{sec:two}
Throughout this section all query costs are one, and the residual instance has at most two uncertain arrivals $f_1,f_2$, in this order, and any number of trivial arrivals. By Lemma~\ref{lem:uncertain} this holds whenever $u(M,I)\le2$, in particular in every block of a graph in $\mathcal C_2$. By Lemma~\ref{lem:account} it suffices to design a level rule with $\waste(Q^*)\le|Q^*|/2$. Trivial arrivals have level one and witness sets inside every certificate; they contribute nothing to $\waste(Q^*)$, and an element outside $Q^*$ receives potential only from $f_1$ and $f_2$.

\paragraph{One arrival.}
If $f$ is the only uncertain arrival with a nonempty witness set and $t_f=1/(2|X_f|)$, then $\waste(Q^*)\le|Q^*|/2$ for every certificate $Q^*$. Indeed, by the witness property either $f\in Q^*$ and $\waste(Q^*)\le|X_f|\,t_f=1/2$, or $X_f\subseteq Q^*$ and $\waste(Q^*)\le1-1/(2|X_f|)\le|X_f|/2$, because $(|X_f|-1)^2\ge0$. We refer to this as the \emph{single-arrival bound}.

\paragraph{Two arrivals.}
The working basis is $A$ when $f_1$ arrives. Let $C_1=C(A,f_1)$ with witness set $X_1$, and let $C_2^0=C(A,f_2)$ be the circuit that $f_2$ closes in the initial basis, with
$X_2^0=\{a\in A\cap C_2^0:a\text{ uncertain},\ U_a>L_{f_2}\}$.
Both are known before any query; if there is no second uncertain arrival, put $X_2^0=\varnothing$. Let $m$ be the canonical maximum deleted at $f_1$. If $m=f_1$, the working basis is still $A$ when $f_2$ arrives, because trivial arrivals do not change it, so $C_{f_2}=C_2^0$ and $X_{f_2}=X_2^0$; we call this outcome \emph{stable}. If $m\ne f_1$, we call the outcome \emph{rerouted}; then $m\in A$ has left the working basis for good, and the circuit of $f_2$ may be arbitrary. In a theta graph with paths $P_1\ni f_1$, $P_2\ni f_2$, and $P_3\subseteq A$, we have $C_1=P_1\cup P_3$ and $C_2^0=P_2\cup P_3$; if an edge of $P_3$ is deleted, $f_2$ closes the cycle $P_1\cup P_2$ instead.

Put $n_1=|X_1\setminus X_2^0|$, $n_{12}=|X_1\cap X_2^0|$, and $n_2=|X_2^0\setminus X_1|$. The \emph{lookahead rule} is as follows.
\begin{enumerate}
\item[\textnormal{(L1)}] If $X_1\ne\varnothing$, set $t_1=1/(2n_1+n_{12})$.
\item[\textnormal{(L2)}] If $X_1=\varnothing$ or the outcome is rerouted, set $t_2=1/(2|X_{f_2}|)$.
\item[\textnormal{(L3)}] If $X_1\ne\varnothing$ and the outcome is stable, set
\begin{equation}\label{eq:look}
 t_2=\max\bigl\{t\in[0,1]:n_{12}(t-t_1)_++n_2\,t\le n_1t_1\bigr\}.
\end{equation}
\end{enumerate}
Rule (L3) is a water-filling step on $X_2^0$ with budget $n_1t_1$, starting from potential $t_1$ on the shared elements. The two budgets $|X_1|t_1$ and $n_1t_1$ add up to exactly one, which is half the cost of the certificate candidate $\{f_1,f_2\}$; a constant budget splits this total evenly, whereas (L1) shifts it to the first step in proportion to the shared elements. For instance, if $X_1=X_2^0$ has two elements, then $t_1=t_2=1/2$, whereas constant budgets produce the levels $\alpha/2$ and $\alpha$. We write \LA\ for Algorithm~\ref{alg:wf} with this rule; the outcome is determined by the realization, so this is a level rule.

\begin{lemma}\label{lem:reroute}
Let $m$ be the canonical maximum of $C_1=C(A,f_1)$ for the first uncertain arrival $f_1$ of a residual instance. If $m\ne f_1$, then $m\in X_1$, and every certificate contains both $f_1$ and $m$.
\end{lemma}
\begin{proof}
Since $m\ne f_1$ is canonical, $w_m>w_{f_1}$ and $w_m\ge w_x$ for all $x\in C_1$. By~\eqref{eq:limits}, $L_m\le L_{f_1}$ and $U_m\le U_{f_1}$. If $m$ were trivial, then $w_m=L_m\le L_{f_1}\le w_{f_1}$; if $f_1$ were trivial, then $w_m\le U_m\le U_{f_1}=w_{f_1}$. Hence both are uncertain, and $U_m>w_m>w_{f_1}>L_{f_1}$ shows $m\in X_1$.

Let $Q$ be a certificate and $T$ a basis that is minimum for every completion consistent with $Q$. For a number $\theta\le L_{f_1}$ let $\Lambda_\theta$ be the set of elements $x\ne f_1$ with $w_x\le\theta$. We claim that every element of $\Lambda_\theta$ is spanned by $\Lambda_\theta\cap A$. An arrival after $f_1$ has weight greater than $L_{f_1}$: this is clear if its lower limit exceeds $L_{f_1}$ or if it is uncertain, and a trivial arrival with lower limit $L_{f_1}$ would precede the uncertain $f_1$. An arrival $g\in\Lambda_\theta$ before $f_1$ is trivial, and by~\eqref{eq:limits} every $x\in C(A,g)\setminus\{g\}$ satisfies $w_x\le U_x\le w_g\le\theta$, which proves the claim.

Suppose $f_1\notin Q$. If $f_1\in T$, consider the completion that changes only the weight of $f_1$ to a value in $(w_m,U_{f_1})$, which is nonempty as $U_{f_1}\ge U_m>w_m$. Then $f_1$ is the unique heaviest element of $C_1$ and lies in no minimum basis, a contradiction. If $f_1\notin T$, every $x\in C(T,f_1)\setminus\{f_1\}$ satisfies $w_x\le w'$ for all $w'\in(L_{f_1},U_{f_1})$, so $f_1$ is spanned by $\Lambda_\theta$ with $\theta=L_{f_1}$, and by the claim also by $\Lambda_\theta\cap A$. As $C_1$ is the unique circuit in $A\cup\{f_1\}$, this gives $C_1\setminus\{f_1\}\subseteq\Lambda_\theta$, and $w_m\le L_{f_1}$ contradicts $w_m>w_{f_1}$.

Suppose $m\notin Q$. If $m\in T$, the completion that changes only the weight of $m$ to a value in $(w_m,U_m)$ makes $m$ the unique heaviest element of $C_1$, a contradiction. If $m\notin T$, every $x\in C(T,m)\setminus\{m\}$ satisfies $w_x\le L_m\le L_{f_1}$; in particular $x\ne f_1$. Thus $m$ is spanned by $\Lambda_\theta$ with $\theta=L_m$, and by the claim also by $\Lambda_\theta\cap A$, which does not contain $m$ because $w_m>L_m$. This contradicts the independence of $A$.
\qed
\end{proof}

\begin{table}[t]
\caption{The cases with $N=n_1+n_{12}+n_2\le3$ in the proof of Theorem~\ref{thm:two}.}\label{tab:small}
\centering\small
\setlength{\tabcolsep}{4.6pt}
\begin{tabular}{@{}lccccccccccccc@{}}
\toprule
$N$ & 1 & \multicolumn{4}{c}{2} & \multicolumn{8}{c}{3}\\
\cmidrule(lr){2-2}\cmidrule(lr){3-6}\cmidrule(l){7-14}
$n_1$ & 0&1&0&1&0&1&0&2&1&0&2&1&0\\
$n_{12}$ & 1&0&1&1&2&0&1&0&1&2&1&2&3\\
$n_2$ & 0&1&1&0&0&2&2&1&1&1&0&0&0\\
\midrule
$t_1$ & $1$&$1/2$&$1$&$1/3$&$1/2$&$1/2$&$1$&$1/4$&$1/3$&$1/2$&$1/5$&$1/4$&$1/3$\\
$t_2$ & $1$&$1/2$&$0$&$2/3$&$1/2$&$1/4$&$0$&$1/2$&$1/3$&$0$&$3/5$&$3/8$&$1/3$\\
\midrule
$2-t_1-t_2$ & $0$&$1$&$1$&$1$&$1$&$5/4$&$1$&$5/4$&$4/3$&$3/2$&$6/5$&$11/8$&$4/3$\\
$N/2$ & $1/2$&$1$&$1$&$1$&$1$&$3/2$&$3/2$&$3/2$&$3/2$&$3/2$&$3/2$&$3/2$&$3/2$\\
\bottomrule
\end{tabular}
\end{table}

\begin{theorem}\label{thm:two}
With unit query costs and at most two uncertain arrivals, \LA\ is $3/2$-competitive.
\end{theorem}
\begin{proof}
Fix a realization and a certificate $Q^*$ of the residual instance; we show $\waste(Q^*)\le|Q^*|/2$. Write $X_2=X_{f_2}$. By the witness property, $Q^*$ contains $f_i$ or all of $X_i$, for $i=1,2$.

\emph{Case A: $X_1=\varnothing$.} By Lemma~\ref{lem:reroute}, $m=f_1$, and among the uncertain arrivals only $f_2$ may have a nonempty witness set. The single-arrival bound applies by (L2).

\emph{Case B: $X_1\ne\varnothing$ and the outcome is rerouted.} By Lemma~\ref{lem:reroute}, $f_1,m\in Q^*$ and $m\in X_1$; moreover $m\notin X_2$ because $m$ is no longer in the working basis. Note that $t_1\le1/|X_1|$, so the elements of $X_1\setminus\{m\}$ have total level less than one. An element of $X_1\cap X_2$ has potential $\max\{t_1,t_2\}\le t_1+t_2$. If $X_2=\varnothing$, then $\waste(Q^*)<1\le|Q^*|/2$. If $f_2\in Q^*$, then
$\waste(Q^*)\le(|X_1|-1)t_1+|X_2|t_2<1+\tfrac12\le|Q^*|/2$,
as $|Q^*|\ge3$. Otherwise $X_2\subseteq Q^*$, the sets $\{f_1,m\}$ and $X_2$ are disjoint, and
$\waste(Q^*)\le(|X_1|-1)t_1+1-t_2<1+|X_2|/2\le|Q^*|/2$,
using $1-1/(2|X_2|)\le|X_2|/2$.

\emph{Case C: $X_1\ne\varnothing$ and the outcome is stable.} Then $X_2=X_2^0$. If $X_2=\varnothing$, then $n_{12}=0$ and $t_1=1/(2|X_1|)$, and the single-arrival bound applies. Assume $n:=n_{12}+n_2\ge1$ and let $B=n_1t_1\le1/2$ be the budget in~\eqref{eq:look}. Shared elements have potential $t_1+(t_2-t_1)_+$. By~\eqref{eq:look},
\begin{equation}\label{eq:mass}
 n_2t_2\le B\quad\text{and}\quad
 \sum_{x\in X_1\cup X_2}y_x\le|X_1|t_1+B=(2n_1+n_{12})t_1=1 .
\end{equation}
We distinguish which of $f_1,f_2$ belong to $Q^*$.

\emph{C1: $f_1,f_2\in Q^*$.} By~\eqref{eq:mass}, $\waste(Q^*)\le1\le|Q^*|/2$.

\emph{C2: $f_1\in Q^*$, $f_2\notin Q^*$.} Then $X_2\subseteq Q^*$, $|Q^*|\ge1+n$, and $\waste(Q^*)\le B+1-t_2$. If $n\ge2$, this is at most $3/2$. If $n=1$, then $t_2\ge B$: for $n_{12}=1$ rule~\eqref{eq:look} gives $t_2=\min\{1,t_1+B\}$, and for $n_2=1$ it gives $t_2=B$. Hence $\waste(Q^*)\le1$.

\emph{C3: $f_1\notin Q^*$, $f_2\in Q^*$.} Then $X_1\subseteq Q^*$, $|Q^*|\ge|X_1|+1$, and by~\eqref{eq:mass}
$\waste(Q^*)\le1-t_1+n_2t_2\le1+(n_1-1)t_1$.
If $n_1\le1$, this is at most $1\le(|X_1|+1)/2$. If $n_1\ge2$, it is less than $3/2\le(|X_1|+1)/2$.

\emph{C4: $f_1,f_2\notin Q^*$.} Then $X_1\cup X_2\subseteq Q^*$, so $|Q^*|\ge N:=n_1+n_{12}+n_2$, and $\waste(Q^*)=2-t_1-t_2$. This is at most $N/2$ if $N\ge4$. For $N\le3$, Table~\ref{tab:small} lists all triples $(n_1,n_{12},n_2)$ with $n_1+n_{12}\ge1$ and $n_{12}+n_2\ge1$ together with the levels from (L1) and~\eqref{eq:look}; in each of them $2-t_1-t_2\le N/2$.

In all cases $\waste(Q^*)\le|Q^*|/2$, and Lemma~\ref{lem:account} gives the ratio $3/2$.
\qed
\end{proof}

\begin{proof}[of Theorem~\ref{thm:opt}]
Process the blocks independently. For a block $H$, choose among $(M(H),I)$ and $(M(H)^*,-I)$ an instance with $u\le2$. By Lemma~\ref{lem:uncertain} its residual instance has at most two uncertain arrivals, and \LA\ is $3/2$-competitive on it by Theorem~\ref{thm:two}. Lemma~\ref{lem:decomp} combines the blocks and transfers the dual guarantee. Every block of a graph in $\mathcal C_2$ satisfies $u(H)\le\mu(H)\le2$. The rule is evaluated in linear time. The lower bound of $3/2$ holds already for a triangle with unit costs~\cite{megow2017randomization}.
\qed
\end{proof}

It suffices that query costs are uniform within each block.

\section{Tightness}\label{app:tight}
We show that the bound of Theorem~\ref{thm:corank} is exact for \WF$(\alpha)$, and that with unit costs no constant budget reaches $3/2$ on theta graphs. We consider $\alpha\in(0,1]$; for $\alpha>1$ the first instance below gives ratio two.

For $h\ge1$, let $\Theta_h$ be the graph with vertices $s,t,v_0,\ldots,v_h$ and the paths $s$--$v_i$--$t$ for $0\le i\le h$, a block of cycle rank $h$. Put $a=sv_0$ and $f_i=sv_i$ for $i\ge1$, let every edge $v_it$ be trivial of weight zero, and let $I_a=(0,10)$ and $I_{f_i}=(2+i/(h+1),20)$. The common limit tree consists of $a$ and the trivial edges, the arrivals are $f_1,\ldots,f_h$ in this order, and in the realizations below the tree never changes and every witness set is $\{a\}$.

\begin{proposition}\label{prop:tightcost}
For every $h\ge1$ and $\alpha\in(0,1]$ there are query costs and realizations on $\Theta_h$ for which \WF$(\alpha)$ has ratio $1+\alpha$ and $1+h/(2\alpha(h+1))$, respectively. Hence~\eqref{eq:budgetbound} is the exact worst-case ratio for $h$ uncertain arrivals, and no constant budget guarantees less than $\rho_h$.
\end{proposition}
\begin{proof}
First let $c_a=1$, $c_{f_i}=1/h$, $w_a=5$, and $w_{f_i}=15$ for all $i$. Every certificate contains every arrival, because $w_a$ lies in its interval, and the arrivals form a certificate since $15\ge U_a$; thus $\OPT=1$. The levels are $t_i=i\alpha/h$. The algorithm queries every arrival, in a core or completion step, and queries $a$ exactly when $\beta\le\alpha$. Its expected cost is $1+\alpha$.

Now let $c_a=1$, $c_{f_i}=1/(\alpha(h+1))$, $w_a=1$, $w_{f_i}=15$ for $i<h$, and $w_{f_h}=5$. The set $\{a\}$ is a certificate, and every certificate contains $a$ since it contains $a$ or $f_h$ and $w_{f_h}<U_a$; thus $\OPT=1$. The levels are $t_i=i/(h+1)$. If $i$ is the first index with $\beta\le t_i$, the algorithm queries $f_1,\ldots,f_{i-1}$, each certified as the maximum of its cycle, and then $a$, which certifies all cycles. If $\beta>t_h$, it queries all arrivals, and the completion step for $f_h$ queries $a$. Thus $a$ is always queried, $f_i$ is queried exactly when $\beta>t_i$, and the expected cost is
\[
 1+\sum_{i=1}^h\frac{1}{\alpha(h+1)}\Bigl(1-\frac{i}{h+1}\Bigr)
 =1+\frac{h}{2\alpha(h+1)}.
\]
The maximum of the two ratios is smallest for $\alpha=\sqrt{h/(2(h+1))}$.
\qed
\end{proof}

\begin{proposition}\label{prop:tight}
With unit query costs, for every $\alpha\in(0,1]$ the worst-case ratio of \WF$(\alpha)$ on theta graphs is at least
$\max\{1+\alpha,\,2-3\alpha/4\}\ge11/7$.
\end{proposition}
\begin{proof}
The ratio $1+\alpha$ is attained on $\Theta_2$ with $I_a=(0,3)$, $I_{f_1}=(1,4)$, a trivial edge $f_2$ of weight $5$, $w_a=2$, and $w_{f_1}=7/2$: as in Example~\ref{ex:triangle}, the optimum queries $f_1$, and the algorithm additionally queries $a$ with probability $t_1=\alpha$.

For the second bound take the theta graph with paths $s$--$u$--$t$, $s$--$v_1$--$t$, and $s$--$v_2$--$t$. Let $a_1=su$, $a_2=ut$, $f_i=v_it$, and let the edges $sv_i$ be trivial of weight zero. Set
$I_{a_1}=(0,10)$, $I_{a_2}=(0,4)$, $I_{f_1}=(2,20)$, and $I_{f_2}=(3,20)$,
with $w_{a_1}=w_{a_2}=1$, $w_{f_1}=5$, and $w_{f_2}=7/2$. The common limit tree contains $a_1$, $a_2$, and the trivial edges; the tree never changes, and both witness sets equal $\{a_1,a_2\}$. The set $\{a_1,a_2\}$ is a certificate. By the witness property, every certificate contains $f_1$ or both $a_1$ and $a_2$, and if it contains $f_1$ it also contains $a_1$, since $w_{f_1}<U_{a_1}$. Therefore $\OPT=2$.

The levels are $\alpha/2$ and $\alpha$. If $\beta\le\alpha/2$, the algorithm queries $a_1,a_2$ and stops. If $\alpha/2<\beta\le\alpha$, it queries $f_1$, the completion step queries $a_1$, which certifies $f_1$ as the maximum because $U_{a_2}\le w_{f_1}$, and the core step for $f_2$ queries $a_2$. If $\beta>\alpha$, it queries $f_1$, $a_1$, and $f_2$, and since $w_{f_2}<U_{a_2}$ the completion step queries $a_2$. The expected cost is
$2\cdot\frac{\alpha}{2}+3\cdot\frac{\alpha}{2}+4(1-\alpha)=4-\frac{3\alpha}{2}$,
giving ratio $2-3\alpha/4$. The two bounds coincide at $\alpha=4/7$.
\qed
\end{proof}

On the last instance, \LA\ uses $t_1=t_2=1/2$ and has expected cost $2\cdot\frac12+4\cdot\frac12=3$, which is $3/2$ times the optimum.

\section{Open Problems}
For unit costs, Theorem~\ref{thm:opt} is tight, and by Proposition~\ref{prop:tight} the ratio $3/2$ cannot be reached with a constant budget. Two questions remain open at the boundary of our results. With arbitrary query costs and two uncertain non-tree edges per block, the best upper bound is $\rho_2=1+1/\sqrt3$ against the lower bound $3/2$. The proof of Theorem~\ref{thm:two} does not extend, because the deleted element $m$ of Lemma~\ref{lem:reroute} may be arbitrarily cheap, so that its presence in every certificate does not pay for the potential spent before rerouting. With unit costs and three uncertain non-tree edges per block, the best upper bound is $\rho_3$. A uniform improvement over $1+1/\sqrt2$ on unrestricted instances remains open as well.

\section*{Acknowledgments}

The motivation for this work originated during an internship at the Max Planck Institute for Informatics. We are especially grateful to Jacob Focke and Philip Wellnitz, who introduced this problem and collaborated on its initial investigation. Their guidance, insights, and discussions were instrumental in shaping the direction of this research.

\section*{AI Disclosure}

During the preparation of this manuscript, Anthropic's Claude Fable 5.2 and OpenAI's ChatGPT Astra 6 were used for checking and reviewing the results, as well as for drafting, editing, and typesetting the text.

\bibliographystyle{splncs04}
\bibliography{main}

@article{erlebach2008computing,
  title={Computing minimum spanning trees with uncertainty},
  author={Erlebach, Thomas and Hoffmann, Michael and Krizanc, Danny and Mihal'{\'A}k, Mat{\'u}s and Raman, Rajeev},
  journal={arXiv preprint arXiv:0802.2855},
  year={2008}
}

@article{megow2017randomization,
  title={Randomization helps computing a minimum spanning tree under uncertainty},
  author={Megow, Nicole and Mei{\ss}ner, Julie and Skutella, Martin},
  journal={SIAM Journal on Computing},
  volume={46},
  number={4},
  pages={1217--1240},
  year={2017},
  publisher={SIAM}
}

@article{focke2020minimum,
  title={Minimum spanning tree under explorable uncertainty in theory and experiments},
  author={Focke, Jacob and Megow, Nicole and Mei{\ss}ner, Julie},
  journal={Journal of Experimental Algorithmics (JEA)},
  volume={25},
  pages={1--20},
  year={2020},
  publisher={ACM, New York, NY, USA}
}

@article{mathwieser2024special,
  title={Special cases of the minimum spanning tree problem under explorable edge and vertex uncertainty},
  author={Mathwieser, Corinna and {\c{C}}ela, Eranda},
  journal={Networks},
  volume={83},
  number={3},
  pages={587--604},
  year={2024},
  publisher={Wiley Online Library}
}

@article{erlebach2022learning,
  title={Learning-augmented query policies for minimum spanning tree with uncertainty},
  author={Erlebach, Thomas and de Lima, Murilo Santos and Megow, Nicole and Schl{\"o}ter, Jens},
  journal={arXiv preprint arXiv:2206.15201},
  year={2022}
}

\end{document}